\documentclass{nature}

\usepackage{amssymb}
\usepackage{amsmath}
\usepackage{amsthm}
\usepackage{graphicx}
\usepackage[nolists,nomarkers]{endfloat}
\usepackage[colorlinks=true,linkcolor=blue,citecolor=blue,pdfauthor={ },pdftitle={ },pdfsubject={ },pdfkeywords={ }]{hyperref}

\usepackage{latexsym}
\usepackage{amssymb}

\usepackage[all]{xy}

\usepackage{subfigure}
\usepackage{dcolumn}
\usepackage{bm}
\usepackage{bbm}
\usepackage{dsfont}
\usepackage{color}
\usepackage{bbold}
\usepackage[usenames,dvipsnames]{xcolor}
\usepackage{amsfonts}
\usepackage{cases}

\graphicspath{{Figures/}}

\title{Interference in Atomic Magnetometry}

\author{Min Jiang$^{1,2}$, Wenjie Xu$^{1,2}$, Qing Li$^{1,2}$, Ze Wu$^{1,2}$, Dieter Suter$^{3}$, \& Xinhua Peng$^{1,2,4}$$^\star$}

\begin{document}

\maketitle
\begin{affiliations}
\item Hefei National Laboratory for Physical Sciences at the Microscale and Department of Modern Physics, University of Science and Technology of China, Hefei 230026, China
\item CAS Key Laboratory of Microscale Magnetic Resonance, University of Science and Technology of China, Hefei, Anhui 230026, China
\item Experimentelle Physik III, Universit$\ddot{\textrm{a}}$t Dortmund, 44221 Dortmund, Germany
\item Synergetic Innovation Center of Quantum Information and Quantum Physics, University of Science and Technology of China, Hefei, Anhui 230026, China
\end{affiliations}

\begin{abstract}
Atomic magnetometers are highly sensitive detectors of magnetic fields that monitor
the evolution of the macroscopic magnetic moment of atomic vapors,
and opening new applications in biological, physical, and chemical science.
However,
the performance of atomic magnetometers is often limited by hidden systematic effects
that may cause misdiagnosis for a variety of applications, e.g., in NMR and in biomagnetism.
In this work, we uncover a hitherto unexplained interference effect in atomic magnetometers,
which causes an important systematic effect to greatly deteriorate the accuracy of measuring magnetic fields.
We present a standard approach to detecting and characterizing the interference effect in, but not limited to, atomic magnetometers.
As applications of our work,
we consider the effect of the interference in NMR structural determination and locating the brain electrophysiological symptom,
and show that it will help to improve the measurement accuracy by taking interference effects into account.
Through our experiments,
we indeed find good agreement between our prediction and the asymmetric amplitudes of resonant lines in ultralow-field NMR spectra -- an effect that has not been understood so far.
We anticipate that our work will stimulate interesting new researches for magnetic interference phenomena in a wide range of magnetometers and their applications.
\end{abstract}


Quantum sensors with new capabilities are revolutionizing methodologies for biological, physical, and chemical science\cite{Degen2017}.
These quantum devices exploit quantum properties or quantum phenomena to detect weak signals,
and can achieve sensitivity and precision approaching the most fundamental limits.
Notable examples include superconducting quantum interference devices\cite{Greenberg1998,Vasyukov2013}, or superconducting qubit sensors\cite{Danilin2018, Shlyakhov2018, Bal2012},
atomic magnetometers\cite{Budker2007, Shah2007}, and nitrogen-vacancy centres\cite{Taylor2008,Zangara2019}.
Their capabilities of high sensitivity and spatial resolution provide new opportunities in applied physics and other areas of frontier science.
For example, nitrogen-vacancy centres have unlocked the door for nanoscale magnetic resonance\cite{Shi2015, Aslam2017, Glenn2018}. Atomic magnetometers have been demonstrated with subfemtotesla sensitivity\cite{Kominis2003, Dang2010, Sheng2013}
and quickly become the promising modality for precision magnetic measurements\cite{Boto2018,Jiwei2018,Xu2006}.
Particularly, atomic magnetometers are applied to detect nuclear magnetic resonance ($\textrm{NMR}$) in ultralow magnetic field,
and deliver new promising applications ranging from chemical analysis\cite{Theis2011, Blanchard2016, Tayler2017,Ledbetter2008q},
quantum control\cite{Bian2017, Jiang2018, Jiang2018v2, Ji2018} to fundamental physics\cite{Teng2018, Wu2019, Garcon2019, Garcon2017}.

Although great progress has been achieved for the development of atomic magnetometers,
further improvement of their performance requires a full understanding of the relevant detection processes.
The detection process of magnetometers can be described by the dynamic evolution of atomic spins, such as atomic Bloch equation\cite{Seltzer2004, Ledbetter2008}
and Liouville equation\cite{Allred2002}.
Quasi-steady-state solution of the dynamics equation is usually adopted in previous works\cite{Seltzer2004, Ledbetter2008,Allred2002}.
However, as we will show in this work,
the quasi-steady approximation results in an important systematic effect of measuring magnetic field
and could cause misdiagnosis for a wide range of magnetometer applications, e.g.,
in NMR\cite{Ledbetter2011, Ledbetter2012, Jiang2019} and in biomagnetism\cite{Boto2018, Xia2006, Zhang2020}.
For example, it has been extensively demonstrated that the NMR signals recorded with atomic magnetometers
uncommonly differ from the expected values with a few tens of percent distortion\cite{Ledbetter2011, Ledbetter2012, Teng2018, Jiang2019, Garcon2019, Wu2019}.
This poses a rather puzzling situation and limits the accuracy of determining the structural information and molar concentration of NMR samples\cite{Doddrell1982}.
Moreover, without the prior knowledge of the hidden systematic effect,
it would reduce the accuracy of atomic magnetometers in locating the magnetic field sources,
such as for diagnosing the brain electrophysiological symptom\cite{Boto2018,Xia2006, Zhang2020} and defects in Li-ion battery cells\cite{Hu2020}.
Therefore, there is a pressing need for a subtle analysis of the relevant physical processes and achieving a precise detection model in atomic magnetometry.

In this work,
we uncover a hitherto unexplained interference effect in atomic magnetometry,
which causes a significant systematic effect to deteriorate the accuracy of measuring magnetic fields.
Unlike the common approach that takes quasi-steady approximation\cite{Seltzer2004, Ledbetter2008,Allred2002,Ledbetter2011, Ledbetter2012, Jiang2019},
our work investigates the dynamic response of atomic magnetometers,
which is the essential origin of the interference effect.
We present a standard approach to detecting and characterizing the interference in atomic magnetometers,
and show that it provides a very precise prediction of the detected signals by taking interference effects into account.
As applications of our work,
we discuss the effect of the interference for a variety of applications, e.g., in NMR and biomagnetism.
Through our ultralow-field NMR experiments,
we demonstrate the interference of NMR signal fields
and provide the first explanation for the hitherto unexplained asymmetric amplitudes of resonant lines in NMR spectra,
significantly improving the accuracy for determining the structure and molar concentration of $\textrm{NMR}$ samples\cite{Ledbetter2011, Ledbetter2012, Jiang2019}.
Moreover, we show that understanding the interference behavior greatly increases the information content of atomic magnetometer signals, e.g.,
the light-shift field of alkali-metal atoms\cite{Mathur1968, Rosatzin1990} and the sign of the Land$\acute{\textrm{e}}$ $g$-factor of nuclear spins.
We anticipate that our work to the first finding of the interference effect in atomic magnetometers
will stimulate interesting new researches for magnetic interference phenomena in a wide range of magnetometers.

\section*{RESULTS}

\begin{figure}[t]  
\centering
	\makeatletter
	\def\@captype{figure}
	\makeatother
	\includegraphics[scale=1.6]{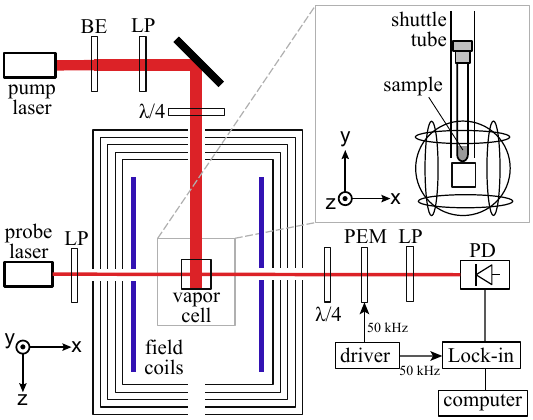} 
	\caption{\textbf{Diagram of experimental setup for an atomic magnetometer and ultralow-field NMR spectrometer}. BE: beam expander, LP: linear polarizer, PD: photodiode, PEM: photoelastic modulator. As shown in the inset, the NMR sample is located at a distance of 1~mm above the Rb vapor cell. See the text and {\color{blue}see Supplemental Information I and Methods} for details.}
	\label{fig0}
\end{figure}

\textbf{Response matrix in atomic magnetometer}. The studied system is an atomic magnetometer (sensitivity $\approx$ $25~\textrm{fT}/\textrm{Hz}^{1/2}$)
with a warm $^{87}$Rb vapor cell ($0.7\times 0.7 \times 1.0$~cm$^3$ and a wall thickness of 1~$\textrm{mm}$), whose setup is described in Fig.~\ref{fig0}.
The vapor cell contains 700~torr of N$_2$ in addition to a small amount of $^{87}$Rb metal,
placed inside a five-layer mu-metal magnetic shield to screen out the ambient magnetic field.
A circularly polarized laser beam (795 nm, 5~mW) optically pumps the $^{87}$Rb atoms along the $z$ direction,
while a linearly polarized probe laser (780 nm, 1~mW) traveling along the $x$ direction provides a Faraday rotation detection signal for the magnetic field.
Moreover, an uniform magnetic field along the $z$ axis is applied to induce a $\textrm{Zeeman}$ effect on $^{87}$Rb atoms,
where the resulting $\textrm{Zeeman}$ precession is much smaller than spin-exchange rate.
More details of the setup are presented in {\color{blue}Methods and Supplemental Information I}.

The electron spin evolution of $^{87}$Rb atoms can be described by a Bloch equation for the electron spin polarization $\textbf{P}$ as follows\cite{Allred2002, Seltzer2004, Ledbetter2008}:
\begin{equation}
\frac{d \textbf{P}}{dt}=\frac{1}{q}[ \mathbf{\Omega}(t) \times  \textbf{P}+R_{\textrm{op}} (\hat{\textbf{z}}-\textbf{P})-\Gamma_{\textrm{SD}} \textbf{P}-\Gamma_{\textrm{pr}} \textbf{P}],
\label{bloch}
\end{equation}
where $\mathbf{\Omega}(t)=g_s \mu_B \textbf{B}(t)$, $g_s\approx 2$ is the electron $\textrm{Land}\acute{\textrm{e}}$ factor,
$\mu_B$ is the Bohr magneton,
$\textbf{B}(t) = [B_{x}(t), B_{y}(t), B_z]$ is the applied magnetic field including to be measured $B_{x}(t)$ and $B_{y}(t)$ along $x$ and $y$ direction and a static bias $B_z$ along $z$ direction,
$q$ is the nuclear slowing-down factor that takes into account the effect of the nuclear spin on the electronic spin\cite{Suter1992},
$R_{\textrm{op}}$ is the optical pumping rate due to the pump laser
and causes spin relaxation because the absorption of a pump beam photon changes the atomic angular momentum,
$\hat{\textbf{z}}$ is the unit vector along the $z$ axis,
$\Gamma_{\textrm{SD}}$ is the relaxation rate due to spin-destruction collisions\cite{Ledbetter2008},
and $\Gamma_{\textrm{pr}}$ is the rate of depolarization due to the probe beam.
Then, the total spin relaxation rate is $\Gamma=R_{\textrm{op}}+\Gamma_{\textrm{SD}}+\Gamma_{\textrm{pr}}$.

For slowly changing magnetic fields $\textbf{B}(t)$, one can obtain the quasi-steady-state solution of Eq.~(\ref{bloch}),
which was used in earlier works\cite{Seltzer2004, Ledbetter2008,Allred2002,Ledbetter2011,Ledbetter2012}.
Unlike the common approach that takes quasi-steady approximation,
we consider a more general case $\textbf{B}(t)=[B_{x0} \cos(2\pi \nu t+ \theta_{x0}), B_{y0} \cos(2\pi \nu t+ \theta_{y0}),B_z]$
where the quasi-steady-state solution might be invalid
and the dynamic response solution must be taken into account.
Although our considered case is simple, it can be applied in a wide range of applications,
from biomagnetic measurement ($\nu\sim 1$-$40$~Hz)\cite{Boto2018, Xia2006} and detection of signals in NMR and magnetic resonance imaging ($\nu\sim 1$-$300$~Hz)\cite{Xu2006,Theis2011, Blanchard2016, Tayler2017,Ledbetter2008q}
to paleomagnetism ($\nu\sim 1$-$100$~Hz)\cite{Dang2010} and searches for axion dark matter ($\nu\sim 1$-$100$~Hz)\cite{Teng2018, Wu2019, Garcon2019, Garcon2017}.
Here $B_{x0}, B_{y0}$ are assumed to be so small that they cause only a perturbation on the atomic polarization \textbf{P}.
Thus, atomic magnetometer can be seen as a linear system and we can describe its response in a form of linear-response theory
({\color{blue}see Methods and Supplemental Information II}):
\begin{equation}
\begin{array}{c}
\left[ {\begin{matrix}
   P_x\\
   P_y \\
\end{matrix}} \right]
= \left[{\begin{matrix}
\Upsilon_{xx}(\nu, B_z) & \Upsilon_{xy}(\nu,B_z)\\
\Upsilon_{yx}(\nu, B_z) & \Upsilon_{yy}(\nu,B_z)\\ \end{matrix}} \right]
 \cdot \left[ {\begin{matrix}
   {\frac{1}{2}B_{x0} {e^{  i (2\pi {\nu }t + {\theta_{x0} })}}}  \\
   {\frac{1}{2}B_{y0} {e^{  i(2\pi {\nu }t  + {\theta_{y0} })}}} \\
\end{matrix}} \right] +\textrm{c.c.}
\end{array}
\label{sig}
\end{equation}
with $\Upsilon_{xx}(\nu,B_z) =  {{\textrm{A}_x}(\nu ,{B_z}){e^{  i{\Phi_x}(\nu ,{B_z})}}} $ and $\Upsilon_{xy}(\nu,B_z) = {{\textrm{A}_y}(\nu ,{B_z}){e^{  i{\Phi_y}(\nu, B_z)}}}$.
The index $x$ and $y$ are permutation symmetric, suggesting $\Upsilon_{yy}(\nu,B_z)=\Upsilon_{xx}(\nu,B_z)$ and $\Upsilon_{yx}(\nu,B_z)=\Upsilon_{xy}(\nu,B_z)$.
The response matrix $\Upsilon(\nu,B_z)$ builds up the relationship between the output signal of the atomic magnetometer and the input magnetic field $[B_{x0} \cos(2\pi \nu t+ \theta_{x0}), B_{y0} \cos(2\pi \nu t+ \theta_{y0})]$.
${\textrm{A}_\xi}(\nu ,{B_z})$ and ${\Phi_\xi}(\nu ,{B_z})$ ($\xi=x,y$) are, respectively, the amplitude and phase response functions, and satisfy the following relations
\begin{eqnarray}
  {\textrm{A}_x}(\nu ,{B_z})/{\textrm{A}_y}(\nu ,{B_z})=\frac{{({{g_s}{\mu _B})}{B_{z}}}}{\sqrt{\Gamma^2+(2\pi \nu)^2 q^2}}  ,\label{eq9a}\\
 \Delta \Phi \equiv {\Phi_x}(\nu ,{B_z})-{\Phi_y}(\nu ,{B_z})=\textrm{arctan}(-\frac{2\pi q \nu}{\Gamma}) . \label{eq9}
\end{eqnarray}

In our experiments, the optical rotation angle of probe laser is proportional to the atomic $P_x$ polarization along the direction of propagation (see Fig.~\ref{fig0}),
thus we only focus on $P_x$ and the partial response matrix $\Upsilon_x(\nu,B_z) \equiv[\Upsilon_{xx}(\nu,B_z), \Upsilon_{xy}(\nu,B_z)]$.
$\Upsilon_x(\nu,B_z)$ is time-independent and only depends on the parameters of the atomic magnetometer.
According to Eq.~(\ref{sig}), it clearly shows that the output signal $P_x$
is a weighted superposition of the $x$ and $y$ components of the magnetic field with the different
amplitude and phase responses if $\textrm{A}_x(\nu ,{B_z}) \neq 0, \textrm{A}_y (\nu ,{B_z}) \neq 0$ and $\Phi_x(\nu ,{B_z})  \neq {\Phi_y}(\nu ,{B_z})$.
For slowly changing magnetic fields (i.e., $\nu$$\to$0),  we reach the quasi-steady-state solution, i.e., $\Phi_x(\nu,B_z)=\Phi_y(\nu,B_z)=0$.
It is also clearly seen, from Eq.~(\ref{eq9a}) and (\ref{eq9}), that the dynamic response solution in ~\eqref{sig} deviates from the quasi-steady-state one when the condition of slowly changing is violated.
As we show below,
this dynamic response feature will bring new experimental phenomena which cannot be explained through the quasi-static assumption.

\begin{figure}[t]  
\centering
	\makeatletter
	\def\@captype{figure}
	\makeatother
	\includegraphics[scale=1.24]{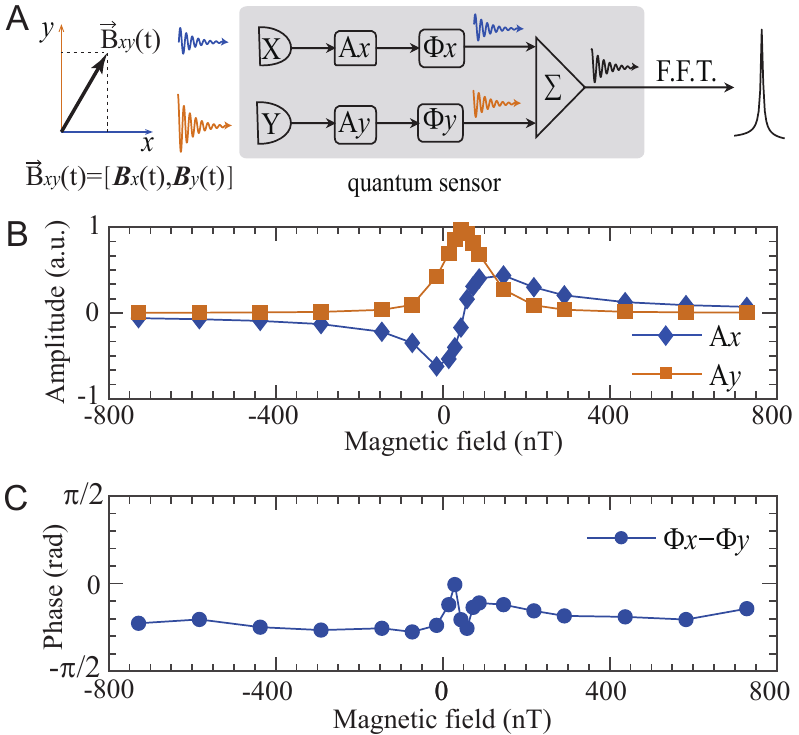} %
	\caption{\textbf{Interference effect in atomic magnetometry and atomic response functions}. (\emph{A}) Basics of the quantum sensor resulting in the interference effect: The magnetic field components along the $x$- and $y$-axes both affect the sensor but the observed signal is given only by the sum of their contributions, which have different sensitivities (amplitudes $\textrm{A}_x$, $\textrm{A}_y$) and phases ($\Phi_x$, $\Phi_y$). (\emph{B}) Experimental amplitude responses $\textrm{A}_x$ and $\textrm{A}_y$ as functions of $B_z$. (\emph{C}) The phase difference between the phase response $\Phi_x(\nu ,{B_z})$ and $\Phi_y(\nu ,{B_z})$ as a function of $B_z$. Here $\nu$ is chosen as 222~Hz for demonstration.
}
	\label{figS3}
\end{figure}

\noindent
\textbf{Interference effect in atomic magnetometer}. Figure~\ref{figS3}\emph{A} schematically shows the basic interference effect resulting from the response of atomic magnetometer
in \eqref{sig}.
The output signal of the atomic magnetometer $s(t)=\alpha P_x(t)$ is an oscillating signal.
Here the proportionality constant $\alpha$ summarises, e.g. amplifier gains and conversion factors of detectors ({\color{blue}see Supplemental Information I}).
The amplitude of the oscillating $s(t)$ signal is
\begin{eqnarray}
 S_{tot}^2 &=& (S_x^2+S_y^2)(1+ \chi \cos \Delta \phi),
 \label{stot}
\end{eqnarray}
where $S_{x}= S_x(\nu ,{B_z}) = \alpha \textrm{A}_x (\nu ,{B_z}) B_{x0}$ and $S_y = S_y(\nu ,{B_z})= \alpha \textrm{A}_y (\nu ,{B_z}) B_{y0}$ respectively denote the amplitude of the output oscillating signal
when there $\textrm{only}$ exists the $x$- or $y$-component in the input magnetic field $\mathbf{B}(t)$.
The term ``$\chi \cos \Delta \phi$'' represents the interference effect when the $x$- and $y$-components are both $\textrm{nonzero}$.
Here the interference phase $\Delta \phi=\Delta \Phi + (\theta_{x0}-\theta_{y0})$ and the interference contrast $ \chi \equiv 2S_x S_y/(S_x^2+S_y^2)$.
Notice that the interference effect vanishes when $S_x=0$ or $S_y=0$ or $\Delta \phi=\pm \pi/2$.
The details of the interference effect depend on both the response matrix and the input magnetic field (or the detected magnetic field).
Importantly, the interference effect results in a systematic error of measuring magnetic fields when the interference is not taken into account.
For example of our atomic magnetometer, when a bias field $B_z\approx 600$~nT is applied
and two oscillating fields are applied with same amplitudes and same initial phases,
$\Delta \Phi \approx \frac{\pi}{5}$ and $|\chi \cos \Delta \Phi|\approx 0.81$ (details are shown in the part of interference calibration) and
accordingly $\sim 5\%$ systematic error of the oscillating field amplitude results from the hidden interference effect.
When this atomic magnetometer is applied to detect NMR signals,
we show below that there is a few tens of percent distortion on NMR peak intensities due to the interference.
Thus, it is worthy to evaluating the interference in atomic magnetometers before they are used as detectors.

Our work provides a standard approach to detecting and characterizing the interference phenomenon in, but not limited to, atomic magnetometers.
Unlike the common approaches that only test the single-axis performance through applying a single-axis calibration field\cite{Allred2002, Ledbetter2008},
we measure the performance of magnetometers under multiple oscillating magnetic fields.
In this way we can obtain hidden dynamic response of magnetometers and extract the important information about the interference effect.
As the response matrix $\Upsilon_x(\nu,B_z)$ is constant for certain $\nu$ and $B_z$,
$\Upsilon_x(\nu,B_z)$ can be experimentally calibrated for a given setup by the following measurements.
An oscillating magnetic field $B_0 \textrm{cos}(2 \pi \nu t)$ is, respectively, applied to the magnetometer along the $\xi$ axis as a reference input,
and then the amplitude $S_{\xi}(\nu ,{B_z})$ of the output signal is recorded as a function of the bias field $B_z$ ({\color{blue}see Supplemental Information II, Fig. S5}).
One can further sweep the frequency $\nu$ of the input oscillating field to obtain $\Upsilon_x(\nu,B_z)$ at arbitrary $\nu$.
Figure~\ref{figS3}\emph{B} shows the experimental measurements for the amplitude responses $\textrm{A}_{\xi}(\nu ,{B_z})$ vs. different bias field $B_z$ at $\nu  = 222$ Hz.
We observe that $\textrm{A}_x(\nu ,{B_z})$ and $\textrm{A}_y(\nu ,{B_z})$ are both non-zero in certain range of $B_z$ around zero.
This illustrates that the atomic magnetometer could be simultaneously sensitive to the magnetic fields along the $x$ and $y$ axes.
The center of symmetry, i.e., the magnetic field where $\textrm{A}_y(\nu ,{B_z})$ reaches its maximum and $\textrm{A}_{x}(\nu,B_z)=0$, is slightly shifted away from zero.
This is because the total magnetic field acting on the $^{87}$Rb atoms is the sum of the external bias field
and the light-shift field $B_{\textrm{LS}}$ due to the virtual transitions of the $^{87}$Rb atoms in the presence of the pump laser\cite{Mathur1968, Rosatzin1990}.
Thus, the response function should take the light-shift field into account by replacing $B_z$ by $\widetilde{B}_z=B_z+ B_{\textrm{LS}}$.

We do not consider the individual phases, but only the difference $\Delta \Phi = \Phi_x(\nu ,{B_z})-\Phi_y(\nu ,{B_z})$,
$\textrm{since}$ the interference effect only depends on the phase difference $\Delta \Phi$, as shown in Eq.~(\ref{stot}).
$\Delta \Phi(\nu ,{B_z})$ was measured from the interference amplitude $S_{{tot}}(\nu ,{B_z})$ of the output signal
when the field $B_0 \textrm{cos}(2 \pi \nu t)$ is simultaneously applied along both $x$ and $y$ axes to the atomic magnetometer
({\color{blue}see Supplemental Information II}).
In this case, $\theta_{x0}=\theta_{y0}=0$.
Using $\textrm{Eq}$.~(\ref{stot}) and the results of $S_x(\nu,B_z)$ and $S_y(\nu,B_z)$ measured above,
$\vert \Delta \Phi(\nu ,{B_z}) \vert$ was experimentally extracted and its sign was further determined by using Eq.~(\ref{eq9}).
Figure~\ref{figS3}\emph{C} shows the experimental data of the phase difference response of the atomic magnetometer.
Obviously, $\Delta \Phi \neq 0 $ illustrates that the phase response of the atomic magnetometer along the $x$ and $y$ axes are different.
More general response matrix and its characterization are presented in {\color{blue}Methods}.

In addition to the above approach of detecting the interference,
we show that it is relatively simple and intuitive to observe the interference effect in practical applications.
This is because the practical magnetic fields, e.g.,
biomagnetic fields\cite{Boto2018, Xia2006, Zhang2020} and NMR fields generated by organic compounds\cite{Ledbetter2011, Ledbetter2012, Jiang2019},
naturally have components along different axes to the magnetometers.
As a typical example, we demonstrate the interference in NMR signals recorded with atomic magnetoemeters
and shall see the interference effect interprets $\textrm{NMR}$ asymmetric spectra, i.e.,
large differences of the amplitudes of resonant lines,
which are greatly distorted from the conventional NMR prediction\cite{Blanchard2016, Tayler2017, Ledbetter2011} and have never been well understood before.

\noindent
\textbf{NMR interference asymmetric spectroscopy}.
Atomic magnetometers, usually operated at ultralow magnetic field,
provide a high-sensitivity detection approach for ultralow-field NMR\cite{Blanchard2016, Tayler2017}.
Our $\textrm{NMR}$ experimental setup is shown in the inset of Fig.~\ref{fig0} and {\color{blue}Supplemental Information I}.
Liquid-state NMR samples are contained in 5-mm NMR tubes,
and pneumatically shuttled between a prepolarizing magnet and a rubidium vapor cell.
The bottom of the NMR tube is $\sim$1 mm above the rubidium vapor cell. During the transfer, a guiding magnetic field ($\approx 1$~G) is applied along the transfer direction to preserve the initial spin state, and is abruptly switched off within 10~$\mu$s prior to signal acquisition.
After the quench, the spins are no longer in an eigenstate of the system Hamiltonian.
The resulting evolution generates a time-dependent magnetic field $\mathbf{B}(t)$ (on the order of pT),
which is detected by the atomic magnetometer.

The equilibrium state ${\rho _{eq}}=1/2^n(\mathbbm{1}+\sum_j \epsilon_j I_{jy})$
(high-temperature approximation, the thermal polarization $\epsilon_j=\gamma_j B_p/k_B T \sim 10^{-6}$) initially prepolarized in a permanent magnet ($B_p\approx 1.3 \ \textrm{T}$) evolves
under the $\textrm{NMR}$ Hamiltonian
$H = \sum_{i;j > i} {2\pi {J_{ij}}} {\textbf{I}_i} \cdot {\textbf{I}_j} - \sum_j {{\gamma _j} B_z  {I_{jz}}  }$,
which generates an oscillating nuclear-spin magnetization $\mathbf{M}(t) = n \hbar \textrm{Tr}[\rho(t) \sum_j \gamma_j \mathbf{I}_j] $ with $\rho (t) = {e^{ - iHt}}{\rho _{eq}}{e^{iHt}} $.
Here ${J_{ij}}$ is the strength of $\textrm{scalar}$ coupling between the $i$th and $j$th spins,
$\mathbf{I}_j=(I_{jx}, I_{jy}, I_{jz})$ represents the $j$th spin angular momentum with gyromagnetic ratio $\gamma _j$,
$n$ is the molecular density of the sample,
and $|\gamma_j B_z |\ll J_{ij}$ for ultralow field NMR. The oscillating nuclear magnetization $\mathbf{M}(t)$ creates $x$ and $y$ oscillating magnetic fields  $\mathbf{B}(t)$ on the atomic magnetometer,
i.e., $[B_x(t), B_y(t)]  \propto [- M_x(t) ,2 M_y(t) ]$ in our case
({\color{blue}see Supplemental Information III}).
The total nuclear magnetization $\mathbf{M}(t)$ in fact comes from the contributions of those in all resonant transition frequencies $\nu _{ab}$, i.e., $\mathbf{M}(t) \equiv n \hbar \sum_{a,b} {\mathbf{M}^{{\nu _{ab}}}(t)}$.
Thus the magnetic field induced by the magnetization has this form $[B_x(t), B_y(t)] \equiv [\sum_{a,b} {B_x^{{\nu _{ab}}}(t)}, \sum_{a,b} {B_y^{{\nu _{ab}}}(t)}] $ with $B_{x}^{\nu _{ab}}(t) = R_{\nu _{ab}} \cos (2 \pi \nu _{ab} t +\theta_{x0}^{\nu _{ab}})$ and $B_y^{\nu _{ab}}(t) = 2 R_{\nu _{ab}} \cos (2\pi \nu _{ab} t +\theta_{y0}^{\nu _{ab}})$.
Here the initial amplitude $R_{\nu _{ab}}$ and phase $\theta_{x0}^{\nu _{ab}}$, $\theta_{y0}^{\nu _{ab}}$ are determined by the initial spin state ${\rho _{eq}}$ and the transition itself at $\nu_{ab}$.
In particular, the phase difference between $\theta_{x0}^{\nu _{ab}}$ and $\theta_{y0}^{\nu _{ab}}$ is useful for calculating the interference term $\chi \cos \Delta \phi$,
i.e., $\theta_{x0}^{\nu _{ab}}-\theta_{y0}^{\nu _{ab}}=\pm \frac{\pi}{2}$,
where the sign $\pm$ depends on ${\rho _{eq}}$ and the transition itself
({\color{blue}see Supplemental Information III}).

\begin{figure}[t]  
\centering
	\makeatletter
	\def\@captype{figure}
	\makeatother
	\includegraphics[scale=1]{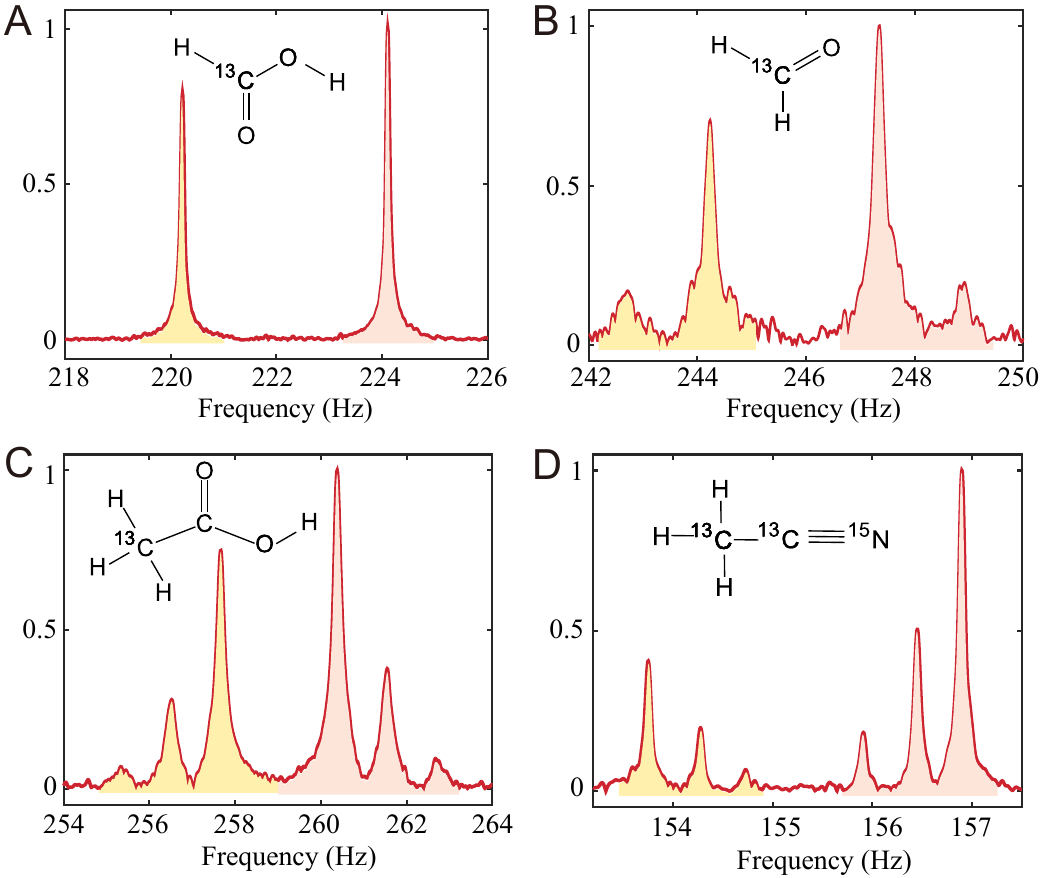} 
	\caption{\textbf{Experimental asymmetric NMR spectra}. (\emph{A}) formic acid, (\emph{B}) formaldehyde, (\emph{C}) acetic acid, (\emph{D}) fully labeled acetonitrile. These are partial spectra of the samples. A small magnetic field ($\approx72.9$~nT) is applied along $z$ in the experiments. Results show that the amplitudes of the $\textrm{NMR}$ peaks with yellow color are different from the corresponding peaks with red color.}
	\label{fig1}
\end{figure}

$\textrm{Figure}$~\ref{fig1} shows the experimentally observed NMR spectra of some samples, clearly exhibiting asymmetric characteristics on their spectral amplitude.
For example, the relative NMR peak amplitudes for acetic acid are theoretically predicted to 1:3:6:6:3:1,
but in practical experiments the ratio is measured to be 1:4.6:12.3:16.4:6.2:1.5 (as shown in Fig.~\ref{fig1}\emph{C}),
which significantly differs from the expected ratio and strictly cannot be the criterion to determine the structural information of the sample.
Similar asymmetric $\textrm{NMR}$ spectroscopy has also been observed in other ultralow-field NMR experiments\cite{Ledbetter2011, Ledbetter2012, Teng2018, Jiang2019, Garcon2019, Wu2019},
in which the asymmetric phenomena are ubiquitous with the detection of atomic magnetometers.
We show that the interference can indeed yield new insights into these asymmetric phenomena.
According to the interference effect in Eq.~(\ref{stot}),
the observed spectral amplitude for each NMR transition peak at $\nu_{ab}$ depends on the response phase difference $\Delta \phi_{\nu_{ab}}=\Delta \Phi(\nu_{ab},B_z) + \theta_{x0}^{\nu_{ab}}-\theta_{y0}^{\nu_{ab}}$ and the $\textrm{contrast}$ $ \chi_{\nu_{ab}} = 4\textrm{A}_x (\nu_{ab},{B_z}) \textrm{A}_y (\nu_{ab},{B_z}) /[\textrm{A}_x^2 (\nu_{ab} ,{B_z})+4 \textrm{A}_y^2 (\nu_{ab},{B_z})]$, which depend on the difference frequency $\nu_{ab}$.
When $\chi_{\nu_{ab}} \cos \Delta \phi_{\nu_{ab}} >0$ NMR peak exhibits constructive interference or when $\chi_{\nu_{ab}} \cos \Delta \phi_{\nu_{ab}} <0$ $\textrm{NMR}$ peak exhibits destructive interference,
which leads to the asymmetric spectra.
The detailed analysis is presented in
{\color{blue}Supplemental Information IV}.
In contrast, the quasi-steady-state solution with the assumption of $\Delta \Phi=0$
presents $\chi_{\nu_{ab}} \cos \Delta \phi_{\nu_{ab}} =0$ and thus fails to explain the asymmetric NMR spectra.

\begin{figure}[t]  
\centering
	\makeatletter
	\def\@captype{figure}
	\makeatother
	\includegraphics[scale=1.12]{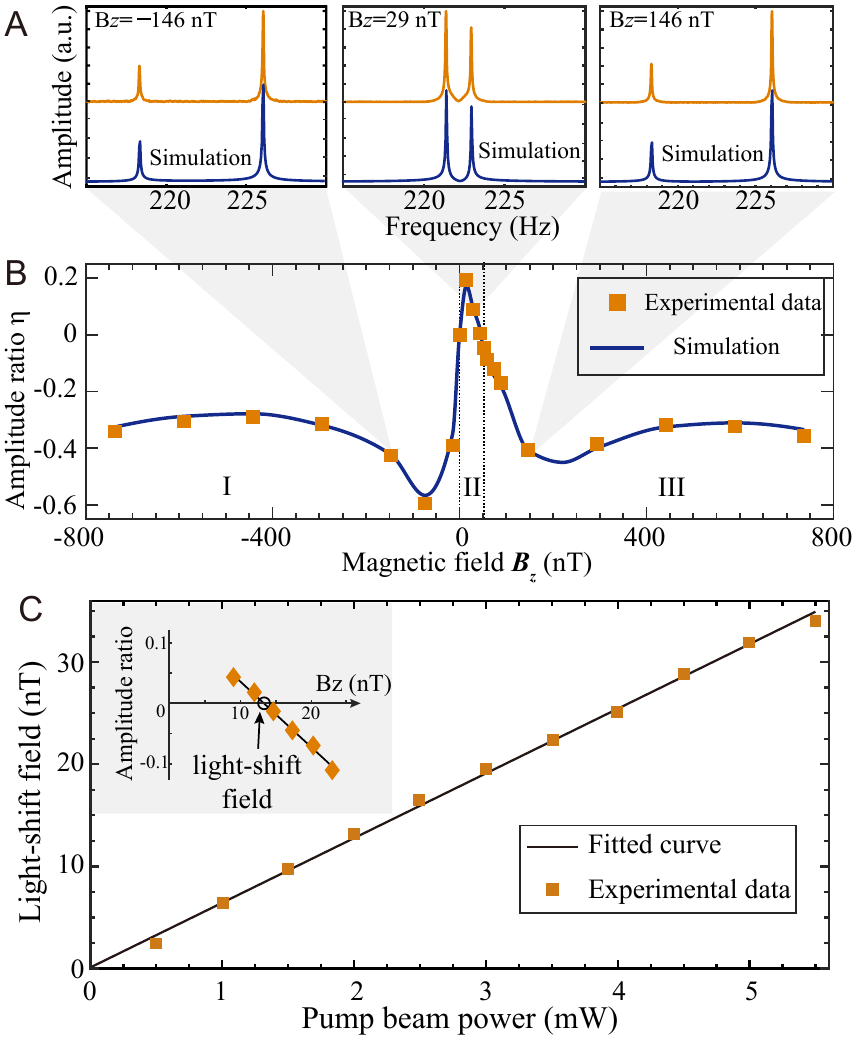} %
	\caption{\textbf{Interference at various magnetic fields and pump beam power}.(\emph{A}) NMR spectra of formic acid at different bias magnetic fields $B_z$. (\emph{B}) Plot of the amplitude ratios of formic acid doublet as a function of fields $B_z$. The experimental data are shown with brown squares, which are in good agrement with our theoretical simulations (blue triangles). (\emph{C}) Plot of light-shift field as a function of pump beam power. The light-shift field is determined by the cross point in the plot of amplitude ratios, as shown in the inset. The light-shift field is directly proportional to the pump beam power.}
	\label{fig3}
\end{figure}

We quantitatively investigate the asymmetry of the NMR spectra at different bias fields $B_z$.
As discussed previously, $\textrm{A}_\xi(\nu ,{B_z})$ and $\Phi_\xi(\nu ,{B_z})$ of the atomic magnetometer vary with the external field ${B_z}$,
and thus the resulting asymmetry of NMR spectra should also depend on ${B_z}$. Take the two-spin system (the formic acid) as the test system.
Figure~\ref{fig3}\emph{A} shows the NMR spectra at the different $B_z$, from which we can clearly see that the asymmetry of the doublet strongly depends on the bias magnetic field $B_z$.
Define the amplitude asymmetry of the formic acid doublet as $\eta \equiv ({\textrm{Amp}_l} - {\textrm{Amp}_h})/({\textrm{Amp}_l} + {\textrm{Amp}_h})$, where ${\textrm{Amp}_l}$ ($\textrm{Amp}_h$) is the amplitude of the $\textrm{NMR}$ peak at lower (higher) frequency.
Figure~\ref{fig3}\emph{B} also shows that the measured asymmetry $\eta$ (brown squares) varies with the change of the bias field $B_z$,
along with the theoretical simulations (blue triangles) obtained by using Eq.~(\ref{sig}) and the experimental response function in Fig.~\ref{figS3}.
In most cases (region I and III in Fig.~\ref{fig3}\emph{B}) the NMR spectrum exhibits negative asymmetry ($\eta_\textrm{I} < 0$ and $\eta_{\textrm{III}} < 0$) while close to the zero bias field (region II in Fig.~\ref{fig3}\emph{B}) $\eta_{\textrm{II}} > 0$.

As observed in Fig.~\ref{fig3}\emph{B},
the amplitude ratio crosses through zero at two different fields.
One point is for $B_z=0$, where the doublet of formic acid collapse into a single resonant peak and thus $\eta$ is defined as zero.
The other point is nontrivial,
resulting from the light-shift field $B_{\textrm{LS}}$
that the $^{87}$Rb atoms experience in the presence of a laser field\cite{Mathur1968, Rosatzin1990}.
Hence the effective field experienced by $^{87}$Rb atoms is $\tilde{B}_{z} = B_z+ B_{\textrm{LS}}$,
which enters the response function $\textrm{A}_x(\nu, \tilde{B}_{z})/\textrm{A}_y(\nu, \tilde{B}_{z}) \propto \tilde{B}_{z}$ in Eq.~(\ref{eq9a}).
When the external field $B_z$ is precisely set to compensate the light-shift field, i.e., $B_z=-B_{\textrm{LS}}$,
the vapor cell equivalently works in zero field and
the amplitude response $\textrm{A}_x(\nu, \tilde{B}_{z})/\textrm{A}_y(\nu, \tilde{B}_{z}) =0$ at any frequency.
In this case, the interference contrast $\chi$ is zero between the $x$ and $y$ magnetic field components.
This importantly suggests a method to eliminate the NMR spectral asymmetry by choosing a proper external bias field or pump beam power.
When NMR signals are measured at such conditions,
the experimental spectra can be recovered to the symmetric spectra as expected by conventional NMR theory.
Our method can be immediately applied to improve the performance of ultralow-field NMR experiments in recent works\cite{Teng2018, Garcon2019, Wu2019}.

We further show that the asymmetric NMR spectra,
which were so far considered to be an unwanted distortion,
can be highly useful for obtaining new information.
Firstly, the spectral asymmetry can be used as an approach to measuring the light-shift field,
which widely exists in atomic magnetometer, atomic clock, and atomic spin gyroscope.
Figure~\ref{fig3}\emph{C} shows the measured light-shift field is directly proportional to the pump beam power,
which is good agreement with the theoretical expectations for the light-shift field.
In contrast with existing approaches\cite{Seltzer2004}, our method does not require additional drive or frequency modulation,
and it is simply based on the spectral asymmetry of NMR sample to determine the light shift.
This unique advantage makes our method particularly well suited to light-shift measurements,
as it can produce no significant static or radio frequency fields contamination.
Secondly,
the NMR interference asymmetric spectroscopy can also provide the sign information of the Land$\acute{\textrm{e}}$ $g$-factor of nuclear spins.
This comes from the important result of the spectral asymmetry: $\eta_{\textrm{I}}/|\eta_{\textrm{I}}|=\eta_{\textrm{III}}/|\eta_{\textrm{III}}|=g/|g|$,
where $g$ is the Land$\acute{\textrm{e}}$ $g$-factor, $\eta_{\textrm{I}}$ and $\eta_{\textrm{III}}$ denotes the asymmetry ratios in region I and $\textrm{III}$, respectively.
For example,
the Land$\acute{\textrm{e}}$ $g$-factor of the $f=1$ manifold in formic acid is negative
due to the asymmetry ratios $\eta_{\textrm{I}},\eta_{\textrm{III}}  <0$ (see Fig.~\ref{fig3}).
The detailed proof and generalized result are presented in {\color{blue}Supplemental Information IV}.
The measurement of the absolute sign of Land$\acute{\textrm{e}}$ $g$-factor overcomes the dilemma in conventional NMR spectroscopy
where the signs of Land$\acute{\textrm{e}}$ $g$-factor are usually lost or only the relative signs can be determined.

\section*{DISCUSSION}
\noindent
Based on the dynamic response of atomic magnetometers,
we build up a comprehensive model to subtly analyze the signals detected by atomic magnetometers.
Furthermore, we uncover a novel interference effect in magnetometers due to the contributions from different detection paths to the observed signal.
Our work could be immediately applied to characterize the interference effect
in applications such as biomagnetic measurements\cite{Hari1993} and battery diagnostics\cite{Hu2020} using atomic magnetometers,
and allows for high-accuracy measurements of magnetic fields.
A promising application is using arrays of high-sensitivity atomic magnetometers to monitor the faint magnetic-field signals from human brain activities\cite{Boto2018, Xia2006, Zhang2020}.
The weak brain magnetic field (on the order of 100~fT) generated by neural currents spatially distributes around the magnetic sensor
and its components along different directions probably interface with each other when recorded with atomic magnetometers,
resulting in the systematic error of the actual brain field amplitudes.
If such interference occurs and without the prior knowledge of interference in magnetometers,
it would reduce the accuracy in locating the brain electrophysiological symptom\cite{Boto2018}.
To make the biomedical diagnosis more accurate,
one can use our present standardized approach to evaluating the interference in atomic magnetometers comprising of MEG in advance,
and further take the observed interference parameters into account for the MEG data post-processing,
enabling an improved accuracy in using the atomic magnetometers in practical clinical applications.

Moreover, the interference effect of atomic magnetometers plays a significant role to influence the intensity of NMR peaks,
that is one characteristic property of NMR spectra and critical for determining the structure of organic compounds, molar concentration of samples and NMR-based quantum information processing (QIP)\cite{Jones2000, Luo2018}.
For example,
the distortionless enhancement by polarization transfer (DEPT)\cite{Doddrell1982} experiments use NMR peak intensities to determine functional groups (e.g., methine, methylene and methyl).
Distance information as well as their connectivity between spins can be well derived from relative NMR peak intensities in spin diffusion or correlation experiments\cite{Ernst1987}.
Besides, NMR is one of the several proposed approaches for QIP,
which relies on NMR peak intensities to encode and readout the information of quantum states\cite{Jones2000, Luo2018} and quantum gates\cite{Jiang2018}.
However, NMR intensities may differ from the theoretical values due to the existence of interference and thus,
this makes experimental conclusions become increasingly uncertain.
To overcome this limitation,
our work provides a solution to characterize the hidden interference effect and takes it into account for the analysis of NMR spectra,
enabling a higher confidence for gaining precise knowledge from NMR spectra.

Although the interference has been identifies in many systems,
such as optical interferometry and atom interferometry\cite{Kasevich1991},
it has not been reported in the area of magnetometry, either experimentally or theoretically.
Our work reports for the first time that the interference can indeed occur in atomic magnetometry
and such interference could result in a significant systematic effect of measuring magnetic fields.
While this work focuses on atomic magnetometers,
our approach to testing and characterizing interference effects could readily be extended to examine myriad quantum sensors,
such as nitrogen-vacancy centres\cite{Shi2015, Aslam2017, Glenn2018},
Bose-Einstein condensate magnetometers\cite{Mssel2014},
single-ion magnetometers\cite{Baumgart2016},
and even electric-field sensors\cite{Dolde2011}.
Identification of the interference effect in such quantum sensors will be important
for future precision measurements,
particularly for magnetic resonance spectroscopy\cite{Shi2015,Aslam2017, Glenn2018,Ledbetter2011, Ledbetter2012, Jiang2019},
biomagnetic medical diagnosis\cite{Boto2018, Xia2006}, as well as for their fundamental applications\cite{Teng2018, Wu2019, Garcon2019, Garcon2017}.

\section*{METHODS}
\noindent
\textbf{Experimental setup}.
The $^{87}$Rb vapor cell works in a stable temperature 180$^{\textrm{o}}$C by resistive heating with using twisted coils and 20-kHz AC current
and is placed inside a five-layer mu-metal magnetic shield (shielding factor of $10^6$).
There are two sets of orthogonal three dimensional coils placed around the vapor cell.
One set of coils are used to apply uniform bias magnetic field (below 1~$\mu \textrm{T}$) along the $z$ axis.
The another set of coils connect with arbitrary wave-function generators
and are used to apply oscillating magnetic fields for measuring magnetometer response functions.
The rubidium atoms in the vapor cell are optically pumped with a circularly polarized laser beam,
whose frequency is tuned close to the center of the buffer-gas broadened and shifted D1 line.
The magnetic field is measured via optical rotation of linearly polarized probe laser beam,
whose frequency is detuned from the D2 transition by about 100~GHz.
To suppress the influence of low-frequency noise,
the polarization of the probe laser beam is modulated by a photoelastic modulator,
and the probe signal is demodulated by a lock-in amplifier.

\noindent
\textbf{Atomic spin polarization}.
Based on Eq.~(\ref{bloch}), one can derive
\begin{equation}
\frac{d P_{\pm}}{dt}=\frac{ \pm i\Omega_z  -\Gamma_{\textrm{rel}}  }{q} P_{\pm} \mp \frac{i \Omega_{\pm} (t)}{q}P_z,
\label{eq5}
\end{equation}
where $\mathbf{\Omega}(t)=g_s \mu_B \textbf{B}(t)$, $P_{\pm}=P_x \pm i P_y$, and $\Omega_{\pm} (t)=\Omega_x (t) \pm i \Omega_{y} (t)$.
When the magnetic fields along the $x$ and $y$ directions are small,
the steady-state $z$ component of the electron spin polarization is $P_{z,0}\approx \frac{R_{\textrm{op}}}{\Gamma}$ and
the $P_z$ can be replaced by $P_{z,0}$ in Eq.~(\ref{eq5}).
After replacing $P_z$ by $P_{z,0}$, the solution of the above equation is
\begin{equation}
P_{+}=e^{-\int \frac{ \Gamma-i \Omega_z} {q} dt} [C_1+ P_{z,0} \int  \frac{- i \Omega_{+} (t)}{q}  e^{\int \frac{ \Gamma-i\Omega_z}{q}dt}dt].
\label{eq6}
\end{equation}
This solution describes the general form of the electron spin polarization $P_x$ and $P_y$.
Eqs.~(\ref{sig}), (\ref{eq9a}), (\ref{eq9}) are derived from Eq.~(\ref{eq6}) when the detailed applied field $\textbf{B}(t) = [B_{x}(t), B_{y}(t), B_z]$ is given.
More details are presented in {\color{blue}Supplemental Information II}.

\noindent
\textbf{Generic approach to measuring atomic response matrix}.
Here we consider the generic approach to characterizing the response matrix $\Upsilon (\nu)$ for arbitrary quantum sensors\cite{Budker2007, Glenn2018, Mssel2014, Baumgart2016, Dolde2011}.
The measured magnetic fields are assumed to be small, and thus quantum sensors can be seen as linear systems.
When there is no prior knowledge of the response performance of sensors,
it should measure the response of sensors to all magnetic fields along the $x$, $y$, $z$ axes.
In this case, the response matrix $\Upsilon (\nu)$ can be represented as a 3$\times$3 matrix,
which builds up the connection between the spin polarization $[P_x, P_y, P_z]$ and
the input magnetic field $[B_{x0} \cos(2\pi \nu t+ \theta_{x0}), B_{y0} \cos(2\pi \nu t+ \theta_{y0}), B_{z0} \cos(2\pi \nu t+ \theta_{z0})]$:
\begin{equation}
\begin{array}{c}
\left[ {\begin{matrix}
   P_x\\
   P_y \\
   P_z\\
\end{matrix}} \right]
= \left[{\begin{matrix}
\Upsilon_{xx}(\nu) & \Upsilon_{xy}(\nu) & \Upsilon_{xz}(\nu)\\
\Upsilon_{yx}(\nu) & \Upsilon_{yy}(\nu) & \Upsilon_{yz}(\nu)\\
\Upsilon_{zx}(\nu) & \Upsilon_{zy}(\nu) & \Upsilon_{zz}(\nu)\\ \end{matrix}} \right]
 \cdot \left[ {\begin{matrix}
   {\frac{1}{2}B_{x0} {e^{  i (2\pi {\nu }t + {\theta_{x0} })}}}  \\
   {\frac{1}{2}B_{y0} {e^{  i(2\pi {\nu }t  + {\theta_{y0} })}}} \\
    {\frac{1}{2}B_{z0} {e^{  i(2\pi {\nu }t  + {\theta_{z0} })}}} \\
\end{matrix}} \right] +\textrm{c.c.},
\end{array}
\label{sig2}
\end{equation}
where $\Upsilon_{ij}(\nu,B_z) =  {{\textrm{A}_{ij}}(\nu){e^{  i{\Phi_{ij}}(\nu)}}} $ ($i,j=x,y,z$).
Notably, Eq.~(\ref{sig2}) is the generalized form of Eq.~(\ref{sig}).
To obtain response matrix $\Upsilon (\nu)$,
it is necessary to experimentally measure a class of amplitude response functions $\textrm{A}_{ij}(\nu)$ and phase responses $\Phi_{ij}(\nu)$.
Similar to the description in the main text, the procedure of measuring response functions is described as following:
(1) For arbitrary $\textrm{A}_{ij}(\nu)$, one applies an oscillating magnetic field $B_0 \textrm{cos}(2 \pi \nu t)$ along the $j$ axis, and then measures the amplitude of the output signal $P_i$;
(2) For arbitrary $\Phi_{ij}(\nu)$,
we do not consider the individual phases, but only the difference, for example, $\Delta \Phi_{ix} = \Phi_{ix}(\nu)-\Phi_{iz}(\nu)$ and $\Delta \Phi_{iy} = \Phi_{iy}(\nu)-\Phi_{iz}(\nu)$
which use $\Phi_{iz}(\nu)$ as the reference phase.
The method to measure such phase differences is the same with that described in the main text.
(3) For arbitrary frequency $\nu$, one can sweep the frequency of the applied oscillating field and repeats the above steps (1) and (2).

\begin{addendum}
\item [Acknowledgements] We thank Teng Wu and Rom$\acute{\textrm{a}}$n Picazo Frutos for useful discussions, and Kang Dai for providing the atomic vapor cells. This work was supported by National Key Research and Development Program of China (Grant No. 2018YFA0306600), National Natural Science Foundation of China (Grants Nos. 11661161018, 11927811), Anhui Initiative in Quantum Information Technologies (Grant No. AHY050000), and USTC Research Funds of the Double First-Class Initiative (Grant No.
YD3540002002).

\item[Author contributions]
M. J. designed research, analyzed the data, and wrote the paper;
W. J. X., Q. L., Z. W. performed experiments;
D. S. proofread and edited the paper;
X. H. P. devised the experimental protocols and wrote the manuscript.
All authors contributed with discussions and to the final form of the manuscript.

\item[Competing Interests] The authors declare that they have no competing financial interests.

\item[Correspondence] Correspondence and requests for materials should be addressed to X. H. P. (xhpeng@ustc.edu.cn).

\item[Additional information]

\end{addendum}

\end{document}